\documentclass[runningheads]{llncs}
\usepackage{graphicx}
\usepackage{booktabs}
\usepackage{multirow}
\usepackage{subfigure}
\usepackage{bigstrut}
\usepackage{color}
\usepackage{amsmath}
\usepackage{amsfonts}
\usepackage{cite}
\usepackage[]{hyperref} 
\usepackage{makecell}
\usepackage{verbatim}
\usepackage{CJKutf8}
\usepackage{url}

\usepackage{tabularx}
\usepackage{fdsymbol}
\usepackage[misc]{ifsym}

\begin{document}
\newsavebox\CBox
\def\textBF#1{\sbox\CBox{#1}\resizebox{\wd\CBox}{\ht\CBox}{\textbf{#1}}}
\begin{sloppypar} 
\def\textBF#1{\sbox\CBox{#1}\resizebox{\wd\CBox}{\ht\CBox}{\textbf{#1}}}

\title{A Pairing Enhancement Approach for Aspect Sentiment Triplet Extraction}
\titlerunning{A Pairing Information Augmentation Approach for ASTE}
%
%

%
\author{Fan Yang \and Mian Zhang \and Gongzhen Hu \and Xiabing Zhou$^{(\textrm{\Letter})}$}
\authorrunning{F. Yang et al.}
%
\institute{School of Computer Science and Technology, Soochow University, Suzhou, China 
\email{\{fyangoct,mzhang2,gzhu\}@stu.suda.edu.cn}, 
\email{zhouxiabing@suda.edu.cn}}

\maketitle              
\begin{CJK*}{UTF8}{gbsn}
\begin{abstract}
Aspect Sentiment Triplet Extraction (ASTE) aims to extract the triplet of an aspect term, an opinion term, and their corresponding sentiment polarity from the review texts.
Due to the complexity of language and the existence of multiple aspect terms and opinion terms in a single sentence, current models often confuse the connections between an aspect term and the opinion term describing it.
To address this issue, we propose a pairing enhancement approach for ASTE, which incorporates contrastive learning during the training stage to inject aspect-opinion pairing knowledge into the triplet extraction model. Experimental results demonstrate that our approach performs well on four ASTE datasets (i.e., 14lap, 14res, 15res and 16res) compared to several related classical and state-of-the-art triplet extraction methods. Moreover, ablation studies conduct an analysis and verify the advantage of contrastive learning over other pairing enhancement approaches.

\end{abstract}

\section{Introduction}
\begin{figure}
\centering
\setlength{\belowcaptionskip}{-4mm}
\centerline{\includegraphics[width=0.8\textwidth]{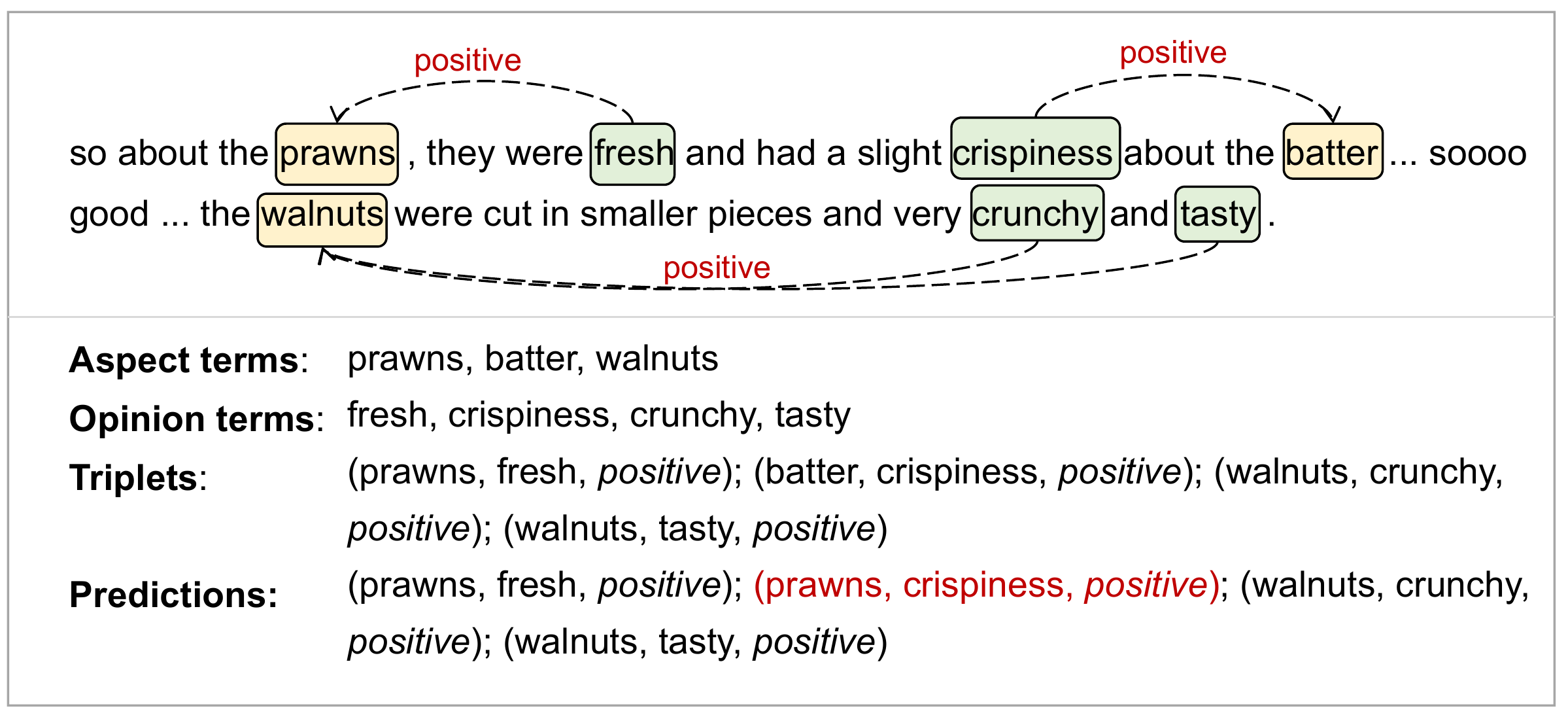}}
\caption{An example of ASTE. The prediction results are generated by \cite{zhang2021towards}.} \label{fig1}
\end{figure}

Aspect-based Sentiment Analysis (ABSA) is an aggregation of several fine-grained sentiment analysis tasks, which involves identifying various aspect-level sentiment elements, including aspect terms, aspect categories, opinion terms, and sentiment polarities\cite{zhang2021aspect}.
Aspect Sentiment Triplet Extraction (ASTE) is a recently proposed subtask of ABSA by Peng et al. \cite{peng2020knowing}, aiming to extract sentiment triplets consisting of an aspect, an opinion, and their corresponding sentiment polarity. Figure~\ref{fig1} shows an example of ASTE.

Given the strong relevance between the subtasks of ABSA, several studies focus on designing unified generation templates and leveraging pre-trained sequence-to-sequence language models to perform multiple tasks simultaneously\cite{zhang2021towards,yan2021unified}. 
Such methods offer a relatively simple way to generate triplets end-to-end without any modifying of the pre-trained models. However, they often fail to match aspects with their corresponding opinions accurately in sentences containing multiple aspects or opinions. This results in poor performance of the triplet extraction task. 
As demonstrated in Figure~\ref{fig1}, the review text contains three aspect terms and four opinion terms. Specifically, the opinion term ``\textit{crispiness}" is used to describe the aspect term ``\textit{batter}". However, since the model does not consider constraining the pairing process when generating triples, it mistakenly matches ``\textit{crispiness}" with ``\textit{prawns}". 

To address this issue, we propose a novel pairing information enhancement approach for ASTE in training stage, where we adopt contrastive learning~\cite{zhang2022optimizing,zhang2022knowledge} to explicitly inject aspect-opinion pairing knowledge into ASTE model. 
Specifically, we first design descriptions about the pairing information between the aspect and opinion terms and encode the descriptions with an independent encoder. 
Then, the dense representations of the aspect-opinion pairs are extracted from the encoding outputs of the triplet extraction model. 
At last, we leverage the contrastive learning objective to push together the matched (mismatched) aspect and opinion and the description of \textit{pair} (\textit{unpair}) in the same vector space, while pushing away the matched (mismatched) against the description of \textit{unpair} (\textit{pair}). 
Experimental results show that our approach performs well on four ASTE datasets (e.g., 14lap, 14res, 15res, 16res) compared to several related classical and state-of-the-art triplet extraction methods.
Moreover, ablation studies further confirm the effectiveness and advantage of our approach in enhancing the pairing ability of the model. 
These findings suggest that it is crucial to improve the pairing ability when aiming to enhance the performance of triplet extraction models, and our approach achieves this effectively.

To summarize, our main contributions include the following:
\begin{itemize}
    \item[$\bullet$] We propose a pairing information enhancement approach for the ASTE task, which has proven to be successful in improving the performance of triplet extraction by enhancing the pairing process.
    \item[$\bullet$] Given the effectiveness of contrastive learning in semantic representation learning, we apply it between the pair type description embeddings and term pair embeddings, which enables the triple extraction models to obtain semantic knowledge related to pair matching. To the best of our knowledge, this is the first work to optimize the pair-matching process for ASTE via contrastive learning.
    \item[$\bullet$] Experimental results on four ASTE datasets show that our methods outperforms the state-of-the-art triplet extraction approaches.
\end{itemize}
\section{Related Work}
\label{related work}
Span-based models are one of the effective methods for extracting triplets\cite{xu2021learning,chen2022span}.
Additionally, machine reading comprehension is also used to solve the triplet extraction problem\cite{chen2021bidirectional}. 
The aforementioned studies are all pipeline-based, which can potentially lead to error propagation. However, joint learning-based approaches can solve this problem well. 
Some researchers design unified annotation schemes, such as the position-aware tagging scheme\cite{xu2020position} and the grid tagging scheme (GTS)\cite{wu2020grid}. 
Other researchers adopt generative frameworks to mine the rich label semantics deeply. 
Mukherjee et al.\cite{mukherjee2021paste} extended the encoder-decoder architecture with a pointer network-based decoding framework. 
Zhang et al.\cite{zhang2021towards} employed the pre-trained T5 model as the generation model, and imported annotation-style and extraction-style modeling paradigms. Han et al.\cite{yan2021unified} exploited BART to solve all ABSA subtasks in an end-to-end framework. 
The approach of employing pre-trained sequence-to-sequence language models to generate triplets is relatively simple, much of the research focuses on designing a unified generative template to satisfy multiple ABSA tasks. 
However, due to the complexity of language and the existence of multiple aspects and opinions in a single sentence, these models often confuse the connections. 
\section{Method}
\label{method}
\begin{figure}[!th]
\centering{\includegraphics[width=\textwidth]{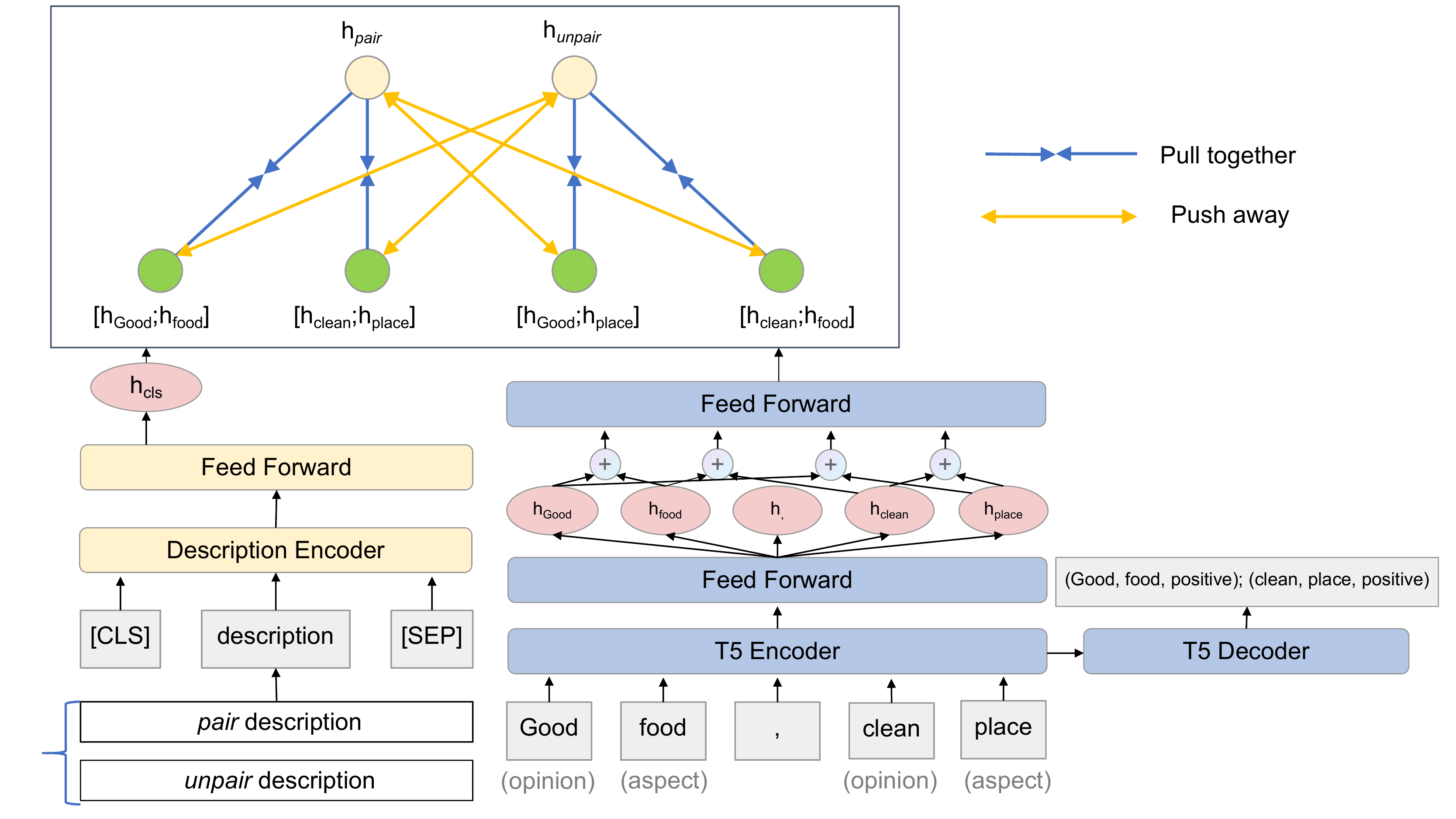}}
\caption{The overall architecture of our approach. 
The description encoder and T5 encoder are used to encode the description of whether a pair is matched and the input text, respectively. 
We use the vector corresponding to \texttt{[CLS]} as the description representation. We apply pooling and concatenation operations to obtain the final term pair representations from the text encoder.
By contrastive learning, we minimize the distance between the description corresponding to \textit{pair} and true term pairs and maximize it with false term pairs.}\label{fig2}
\end{figure}

Our new approach that leverages contrastive learning to explicitly inject pairing knowledge into the triplet extraction model includes three main steps:
(1) we extract the dense representations of the terms from the encoding outputs of the triplet extraction model;
(2) we design descriptions that indicate whether an aspect term and opinion term match and encode them with an independent encoder;
(3) based on the description embeddings and pair embeddings, we adopt contrastive learning objective to enhance the ability of the model to pair aspect and opinion terms.
Figure~\ref{fig2} provides an overview of the proposed approach.

\subsection{Task Formulation}
Given an input sentence $X=\{{x_1},...,{x_n}\}$ with $\mathnormal{n}$ words, the goal of ASTE model is to extract a set of triplets 
$T=\{(a, o, s)_i\}^{\mid T \mid}_{i=1}$  
from $X$, where $a$ and $o$ denote aspect term and opinion term, respectively. The sentiment polarity $s$ of the given aspect can be \textit{positive}, \textit{neutral} or \textit{negative}.

\subsection{Triplet Extraction}
Following the framework proposed by Zhang et al.\cite{zhang2021towards}, we leverage the T5~\cite{raffel2020exploring}, an encoder-decoder-based model pre-trained on a multi-task mixture of unsupervised and supervised tasks, to formulate the triplet extraction task in a generative manner.
Traditional methods use task-specific classification networks to model and predict sentiment terms and tendencies. These methods make predictions in a discriminative manner and use the class indices as labels, which often ignores the rich semantics within the labels.
In contrast, generative methods decode a single sequence that contains triplets, including aspects, opinions, and sentiment polarities, enabling a deeper exploration of the semantics, thus providing a more effective solution to the triplet extraction problem.

\noindent\textbf{Encoder}
The T5 encoder is used to encode the input review text $X$ into the hidden representation $\textbf{H}^{e}$, which contains the semantics of $X$:
\begin{equation}
    \textbf{H}^{e} = \textrm{T5Encoder}({x_1},...,{x_n}),
\end{equation}
where $\textbf{H}^{e} \in \mathbb{R}^{n \times d}$, and $\mathnormal{d}$ is the hidden dimension.

\noindent\textbf{Decoder}
Following Zhang et al.\cite{zhang2021towards}, the output paradigm is divided into annotation style and extraction style. For annotation style, each aspect term is annotated with its corresponding opinion terms and sentiment polarity, i.e., [aspect $\mid$ opinion $\mid$ sentiment polarity]. For extraction style, the target is the triplets concatenation outputs. An example is shown below:
\begin{itemize} 
    \item[$\bullet$] \textbf{Input}: Nice keyboard, battery and screen work ok. 
    \item[$\bullet$] \textbf{Target (Annotation-style)}: Nice [keyboard $\mid$ positive $\mid$ Nice], [battery $\mid$ neutral $\mid$ ok] and [screen $\mid$ neutral $\mid$ ok] work ok.
    \item[$\bullet$] \textbf{Target (Extraction-style)}: (keyborad, Nice, postive); (battery, ok, neutral); (screen, ok, neutral).
\end{itemize}

To generate the outputs defined above, the T5 decoder takes the encoder outputs $\textbf{H}^{e}$ and the previous decoder outputs $Y_{<t}$ as inputs to get current decoder output $\textbf{h}_t^d$. Then, $\textbf{h}_t^d$ is passed through a fully connected softmax layer and mapped to the vocabulary distribution:

\begin{equation}
    \textbf{h}_t^d = \textrm{T5Decoder}(\textbf{H}^{e},Y_{<t}),
\end{equation}
\begin{equation}
    P_t = \textrm{softmax}(\mathbf{W}_v\textbf{h}_t^d+\mathbf{b}_v),
\end{equation}
where $\textbf{h}_t^d \in \mathbb{R}^{d}$, $\mathbf{W}_v$ and $\mathbf{b}_v$ are learnable parameters.

\noindent\textbf{Triplet Generation Objective}
We utilize the cross-entropy function to calculate the loss:
\begin{equation}
\mathcal{L}_{e}=-\sum^n_i y_i\log(P_i)+(1-y_i)log(1-P_i),
\end{equation}
where $y_i$ is the ground truth label, representing the aspect term, opinion term, sentiment label or other component in the target output.

\subsection{Pair Contrastive Learning}

\noindent\textbf{Description Embeddings}
We define $\mathcal{D}=\{D_{pair},D_{unpair}\}$ 
as the set of descriptions indicating whether the aspect terms and opinion terms match, derived from prototypical instances. Given a description $D_k$ with $\mathnormal{m}$ words, we employ BERT~\cite{devlin2019bert} to learn the sequence representation:
\begin{equation}
    \mathbf{h}^{D_k}_{\texttt{[CLS]}} = \textrm{BERT} (\{d_1,...,d_{m}\}),
\end{equation}
where $\mathbf{h}_{\texttt{[CLS]}}$ is the output of BERT corresponding to $\texttt{[CLS]}$. Then, a linear layer is used to obtain the final representation for the description $D_k$.
\begin{equation}
    \mathbf{d}_k =\mathbf{W_d}\mathbf{h}^{D_k}_{\texttt{[CLS]}} + \mathbf{b_d},
\end{equation}
where $\mathbf{d}_k\in \mathbb{R}^{d}$, $\mathbf{W_d}$ and $\mathbf{b_d}$ are learnable parameters.

\noindent\textbf{Term Pair Embeddings}
To generate positive and negative pair instances, according to the position index $I \in \{0,...,n\}$ of aspects and opinions in the input  $\boldsymbol{X}$, we first extract the dense representations of the aspect terms $\mathbf{h}_a$ and opinion terms $\mathbf{h}_o$ from the encoding outputs $\textbf{H}^{e}$:
\begin{equation}
    \mathbf{h}_a = f(\textbf{H}^{e}[I_{a_{start}}:I_{a_{end}}]),
\end{equation}
\begin{equation}
    \mathbf{h}_o = f(\textbf{H}^{e}[I_{o_{start}}:I_{o_{end}}]),
\end{equation}
where $\mathbf{h}_a, \mathbf{h}_o \in \mathbb{R}^{d}$ is the dense representation of aspect terms and opinion terms. f: $\mathbb{R}^{d\times n} \rightarrow \mathbb{R}^{d \times 1}$ is a average pooling function that maps n output vectors to 1 representative vector. 
Then, we perform a linear function on the concatenation of $\mathbf{h}_a$ and $\mathbf{h}_o$:
\begin{equation}
    \mathbf{h}_{c} =\mathbf{W_s}(\mathbf{h}_a \oplus \mathbf{h}_o)+ \mathbf{b_s},
\end{equation}
where $\mathbf{h}_{c} \in \mathbb{R}^{d}$ is the final representation of term pairs, $\mathbf{W_s}$ and $\mathbf{b_s}$ are learnable parameters. 

\begin{figure}
\centering{\includegraphics[width=0.4\textwidth]{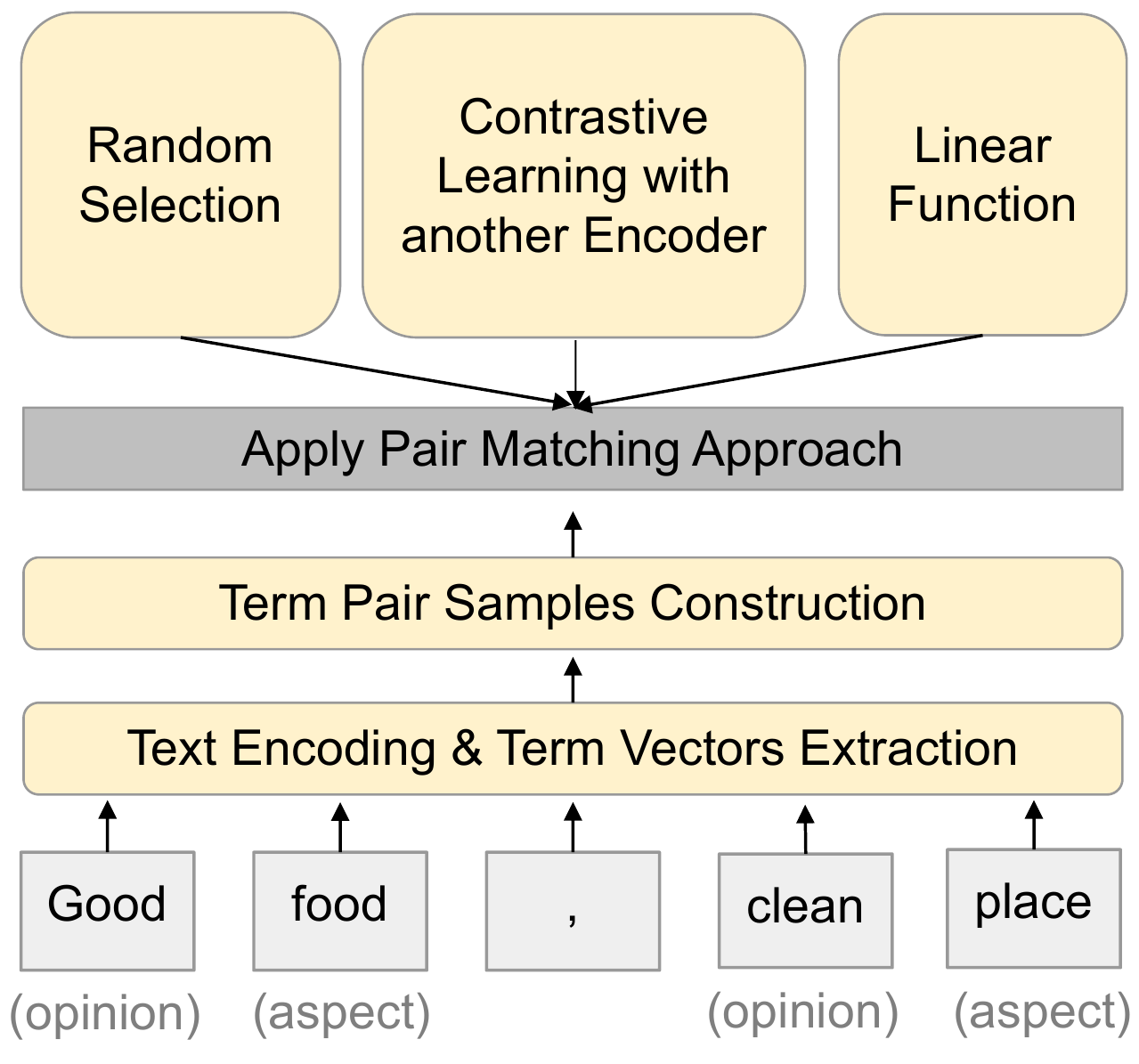}}
\caption{Application of different pairing information enhancement approaches on the triplet extraction model.}\label{fig3}
\end{figure}

\noindent\textbf{Contrastive Learning Objective}
To implement pairing information enhancement, we test three approaches to match the known aspect terms and opinion terms, namely random selection, linear function and contrastive learning. The framework is shown in Figure~\ref{fig3}. The experimental results are presented in Section~\ref{4.6}. 
In the end, we chose contrastive learning with the best performance.

During contrastive learning, we force the representation of \textit{pair} and \textit{unpair} descriptions $d_k$ to be similar with the positive aspect-opinion term pairs, and to be dissimilar with the negative term pairs：
\begin{equation}
   \mathcal{L}_{c}=-log\frac{e^{sim(\mathbf{h}_{c},\mathbf{d}_k) / \tau}}{e^{sim(\mathbf{h}_{c},\mathbf{d}_k) / \tau}+\sum_{\mathbf{h}^{\prime}\in \textbf{H}^-_k }e^{sim(\mathbf{h}^{\prime},\mathbf{d}_k) / \tau}},
\end{equation}
where the pair corresponding to $\mathbf{h}_{c}$ belongs to pair type $D_k$, $\mathbf{h}^{\prime}$ is the set of all negative pairs, $\mathbf{d}_k$ is the pair type description, and $\tau$ is the temperature coefficient.
Since the desriptions represent the central point of \textit{pair} (\textit{unpair}), they contain rich pairing-related semantic information. By employing contrastive learning, the pair composed of aspect and opinion extracted by T5 can learn this information, thus facilitating the extraction of valuable features for pairing. Consequently, this enables the generation of more accurate triplets.
\subsection{Training}
Finally, we combine the above loss functions to form the loss objective of our approach:
\begin{equation}
    \mathcal{L} =\alpha\mathcal{L}_{e}+\beta\mathcal{L}_{c},
\end{equation}
where $\alpha$ and $\beta$ are scalar hyperparameters.
\section{Experiments}
\label{exp}
\subsection{Experimental Setup}
We evaluate our approach on ASTE-Data-V2 released by Xu et al. \cite{xu2020position}, which includes three datasets of restaurant domain (i.e., 14res, 15res, and 16res) and one dataset of laptop domain (i.e., 14lap). 
These datasets originally come from SemEval Challenges \cite{kirange2014aspect,pontiki2015semeval,pontiki2016semeval}. 
The statistics are shown in Table~\ref{tab1}, each dataset contains numerous examples of multi-aspect terms, multi-opinion terms, as well as combinations of both multi-aspect and multi-opinion terms.

\begin{table}
  \centering
  \setlength\tabcolsep{2.5pt}
  \caption{The statistics of ASTE-Data-V2. ($\square$, $\circ$, $\heartsuit$, and $\Diamond$ denote the number of sentences, triplets containing more than one aspect, triplets containing more than one opinion, and  triplets containing more than one aspect and opinion, respectively.)} 
    \begin{tabular}{ccccccccccccccccc}
\toprule[1pt]
    \multirow{2}[4]{*}{Dataset} & \multicolumn{4}{c}{14lap}     & \multicolumn{4}{c}{14res}     & \multicolumn{4}{c}{15res}     & \multicolumn{4}{c}{16res} \bigstrut\\
\cmidrule(lr){2-5}\cmidrule(lr){6-9}\cmidrule(lr){10-13}\cmidrule(lr){14-17}
    & $\square$   & $\circ$   & $\heartsuit$   & $\Diamond$   & $\square$   & $\circ$   & $\heartsuit$   & $\Diamond$   & $\square$   & $\circ$   & $\heartsuit$   & $\Diamond$   & $\square$   & $\circ$   & $\heartsuit$   & $\Diamond$ \bigstrut\\
    \hline
    train &906	&265	&274	&178	&1266	&533	&557	&429	&605	&183	&239	&155	&857	&244	&319	&210  \bigstrut[t]\\
    val   &219	&59	&69	&42	&310	&123	&132	&98	&148	&49	&64	&41	&210	&65	&77	&50 \\
    test &328	&103	&111	&70	&492	&228	&245	&187	&322	&82	&98	&68	&326	&91	&119	&76 \bigstrut[b]\\
 \bottomrule[1pt]
    \end{tabular}%
  \label{tab1}%
\end{table}%

\subsection{Implementation Details}
We implement our approach based on GAS\cite{zhang2021towards} repository and follow their work to choose T5\cite{raffel2020exploring} as the pre-trained language model.
We employ the prediction normalization strategy presented by GAS. 
The description encoder is initialized using BERT-base-cased\footnote{https://huggingface.co/bert-base-cased}.
The output size of the linear layer is set to 128, and the initial temperature parameter is 0.07.
Our model is trained using the AdamW optimizer with a learning rate of 3e-4. We apply a dropout rate of 0.5 and a batch size of 16, and the model is trained up to 20 epochs.
The weights for the loss function are set to $\alpha$=0.9 and $\beta$=0.1. 
We use F1 scores to evaluate the triplet extraction performance.
We report the average score of five runs.

\subsection{Baselines}
We compare the performance of our approach with the following baselines: 
\begin{itemize}
    \item \textbf{Peng-two-stage}\cite{peng2020knowing}, a two-stage framework, which extracted aspect-sentiment pairs and opinion terms in the first stage, followed by Cartesian product pairing and classification in the second stage.
    \item \textbf{GTS-BERT}\cite{wu2020grid}, a unified grid tagging scheme, which designed an inference strategy to exploit mutual indications between different opinion factors.
    \item \textbf{$\mathbf{S}^3\mathbf{E}^2$}\cite{chen2021semantic}, a model based on vanilla GTS and represented semantic and syntactic relations between word pairs by a graph neural network to enhance the triplet extraction performance.
    \item \textbf{BMRC}\cite{chen2021bidirectional}, a method that converted the ASTE task into a multi-turn machine reading comprehension (MRC) task with well-designed queries.
    \item \textbf{ASTE-RL}\cite{yu2021aspect}, a model that treated the aspects and opinions as arguments of sentiment in a hierarchical reinforcement learning framework.
    \item \textbf{BART-ABSA}\cite{yan2021unified}, a generative method that converted the ASTE task into the index generation problem through the BART model.
    \item \textbf{GAS}\cite{zhang2021towards}, a method that employed the T5 model as the generation model, and imported annotation style and extraction style modeling paradigms.
    
\end{itemize}

\subsection{Main Results}
The experimental results on ASTE-Data-V2 are shown in Table~\ref{tab2}. To make a fair comparison with GAS\cite{zhang2021towards}, we implement our approach on both extraction style and annotation style output templates. 
It is worth noting that our approach outperforms other studies among all the baselines considered.
On the 14lap, 14res, 15res, and 16res datasets in extraction style, our approach achieves an F1 score increase of 0.90\%, 0.37\%, 0.68\%, and 1.28\%, respectively, compared to GAS. 
Similarly, for annotation style, we also increase the F1 scores by 7.19\%, 3.17\%, 2.52\%, and 2.29\%, respectively. 
These results indicate that enhancing pairing information is essential for solving the ASTE task, while the training process of vanilla T5 does not consider controlling aspect-opinion pairing, introducing contrastive learning enables the model to learn such information.

\begin{table}[ht]
  \caption{Benchmark evaluation results on ASTE-Data-V2 (F1-score, \%). All baseline results are from the original papers.}
\begin{tabular}{@{}cp{1cm}p{1cm}p{1cm}p{1cm}@{}}
\toprule
\multicolumn{1}{l}{}  & \multicolumn{1}{l}{14lap} & \multicolumn{1}{l}{14res} & \multicolumn{1}{l}{15res} & \multicolumn{1}{l}{16res} \\ \midrule
Peng-two-stage& 42.87& 51.46 & 52.32  & 54.21 \\
GTS-BERT & 54.36& 67.50 & 60.15  & 67.93 \\
$\mathbf{S}^3\mathbf{E}^2$ & 52.01& 66.74 & 58.66  & 66.87 \\
BMRC& 58.18& 68.64 & 58.79  & 67.35 \\
ASTE-RL & 59.50& 69.61 & 62.72  & 68.42 \\
BART-ABSA & 58.69& 65.25 & 59.26  & 67.62 \\
GAS-extraction   & 60.78                   & 72.16                   & 62.10                   & 70.10                   \\
Ours-extraction & \textBF{61.68}          & \textBF{72.53}          & 62.78                   & \textBF{71.38}          \\
GAS-annotation   & 54.31                   & 69.30                   & 61.02                   & 68.65                   \\
Ours-annotation & 61.50                   & 72.47                   & \textBF{63.54}          & 70.94                   \\ 
\bottomrule
\end{tabular}
  \label{tab2}%
\centering
\end{table}

\subsection{Effectiveness of Pairing Information Enhancement}
To prove that the improved performance of triplet extraction shown in Table~\ref{tab2} is indeed due to the enhancement of pairing tasks, that is, contrastive learning is effective for improving the pairing accuracy, we conduct experiments on the Aspect Opinion Pair Extraction (AOPE) task. 
Table~\ref{tab3} displays the results with and without employing contrastive learning.
The table illustrates that the pairing performance has significantly improved with the introduction of contrastive learning, regardless of the annotation style or extraction style. 
This highlights the importance of implementing pairing information enhancement.
\begin{table}[ht]
  \caption{Ablation study on Aspect Opinion Pair Extraction (AOPE), ``CL" represents contrastive learning (F1-score, \%).}
\begin{tabular}{@{}cp{1cm}p{1cm}p{1cm}p{1cm}@{}}
\toprule
                      & 14lap            & 14res            & 15res            & 16res            \\ \midrule
Ours-extraction & \textBF{69.05}  & \textBF{74.74}          & \textBF{67.27}          & 74.50          \\
w/o CL   & 68.08          & 74.12          & 67.19          & \textBF{74.54}          \\ \hline
Ours-annotation & \textBF{69.84} & \textBF{75.50} & \textBF{71.22} & \textBF{76.40} \\
w/o CL  & 69.55          & 75.15          & 67.93          & 75.42          \\
 \bottomrule
\end{tabular}
  \label{tab3}%
\centering
\end{table}

\subsection{Comparison of Pairing Strategies}
\label{4.6}
To demonstrate the advantage of contrastive learning in enhancing pairing information, we conduct a comparison with random selection and linear function-based pairing methods, the comparison framework is illustrated in the Figure~\ref{fig3}. 
In order to increase the controllability of the results and minimize the interference caused by extraction results on the final metrics, we assume that the aspects and opinions in the input sequence are already known, and only focus on evaluating the effectiveness of pairing.
As shown in Table~\ref{tab4}, contrastive learning outperforms the other two methods. 
Random selection performs worst because it does not introduce any prior knowledge. 
In addition, method based on linear function shows an significant improvement over random pairing, but still not as good as contrastive learning, demonstrating the superiority of contrastive learning in learning the information of whether to pair.

\begin{table}[ht]
  \caption{Comparison results of different pairing strategies(F1-score, \%).}
\begin{tabular}{@{}cp{1cm}p{1cm}p{1cm}p{1cm}@{}}
\toprule
    & 14lap            & 14res            & 15res            & 16res            \\ \midrule
random selection           & 47.52          & 49.71          & 47.69          & 48.52                   \\
linear function         & 81.93              & 89.27          & 89.80          & 90.82                    \\
contrastive learning & \textBF{84.18}& \textBF{90.51} & \textBF{91.42} & \textBF{91.49}  \\ \bottomrule
\end{tabular}
  \label{tab4}%
\centering
\end{table}

\subsection{Pair Feature Visualization}
\begin{figure}[ht]
\setlength{\belowcaptionskip}{-1cm} 
\centering
\subfigure[14lap-before]{\includegraphics[width=0.244\textwidth]{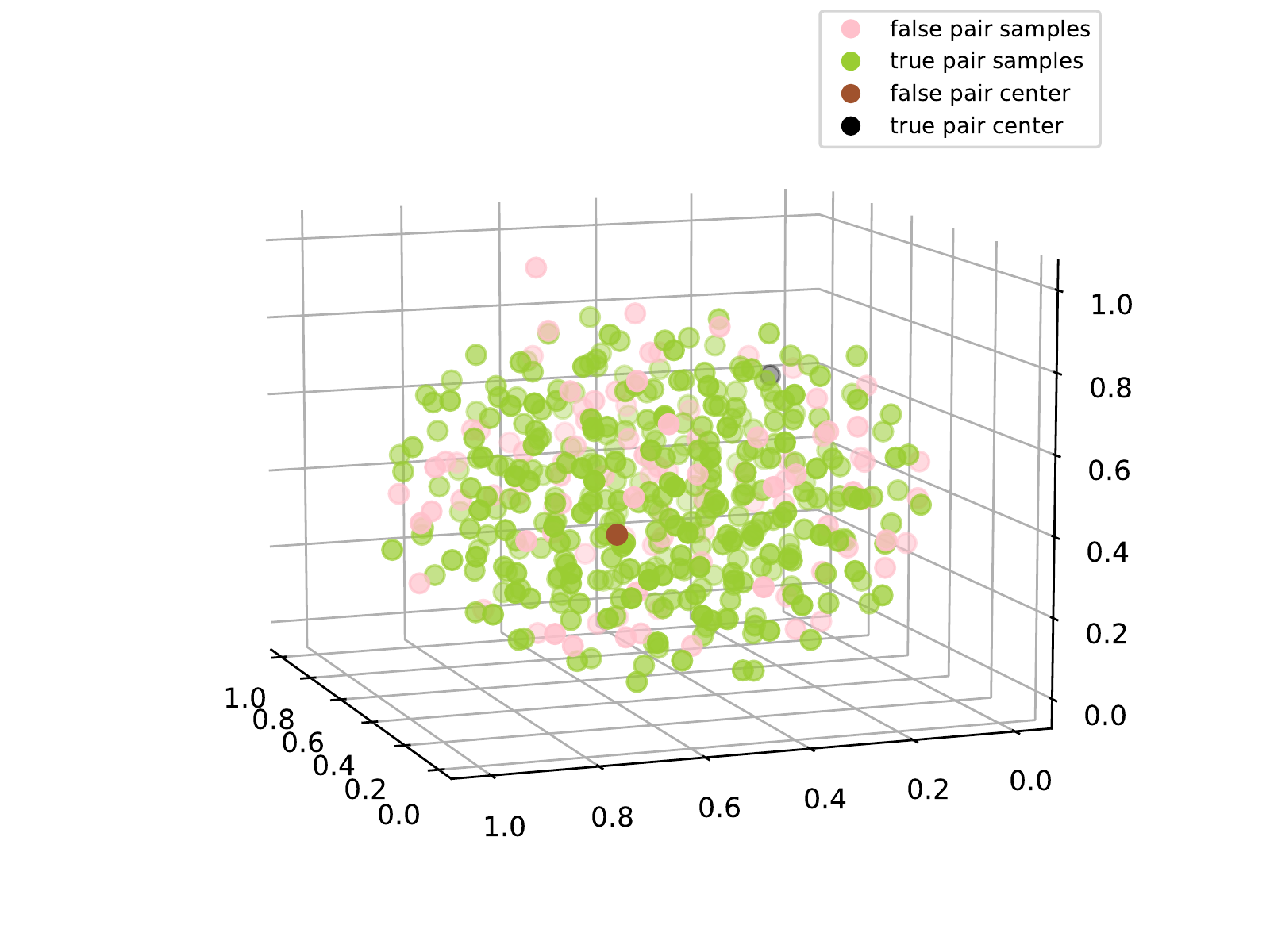}}
\subfigure[14res-before]{\includegraphics[width=0.244\textwidth]{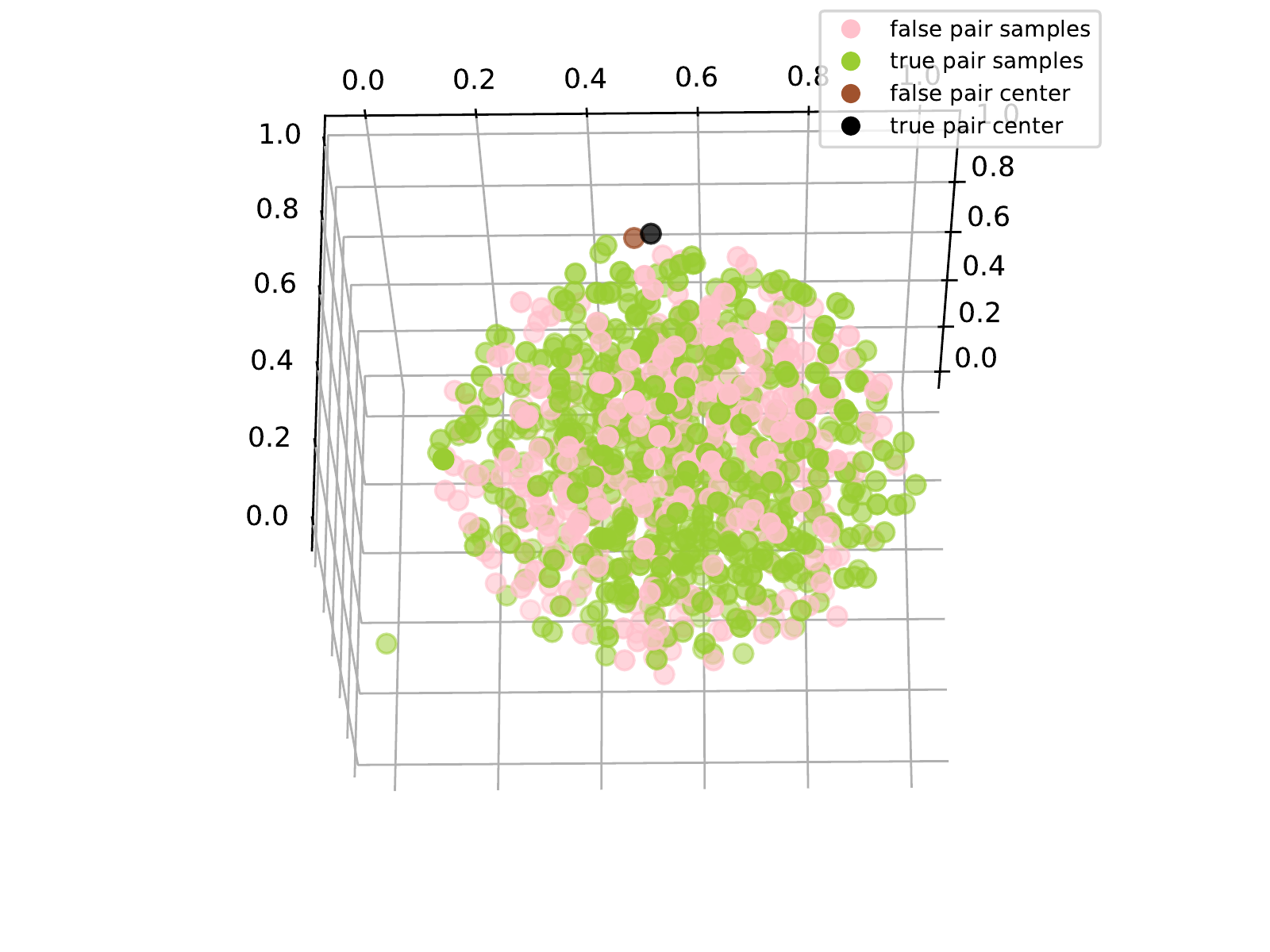}}
\subfigure[15res-before]{\includegraphics[width=0.244\textwidth]{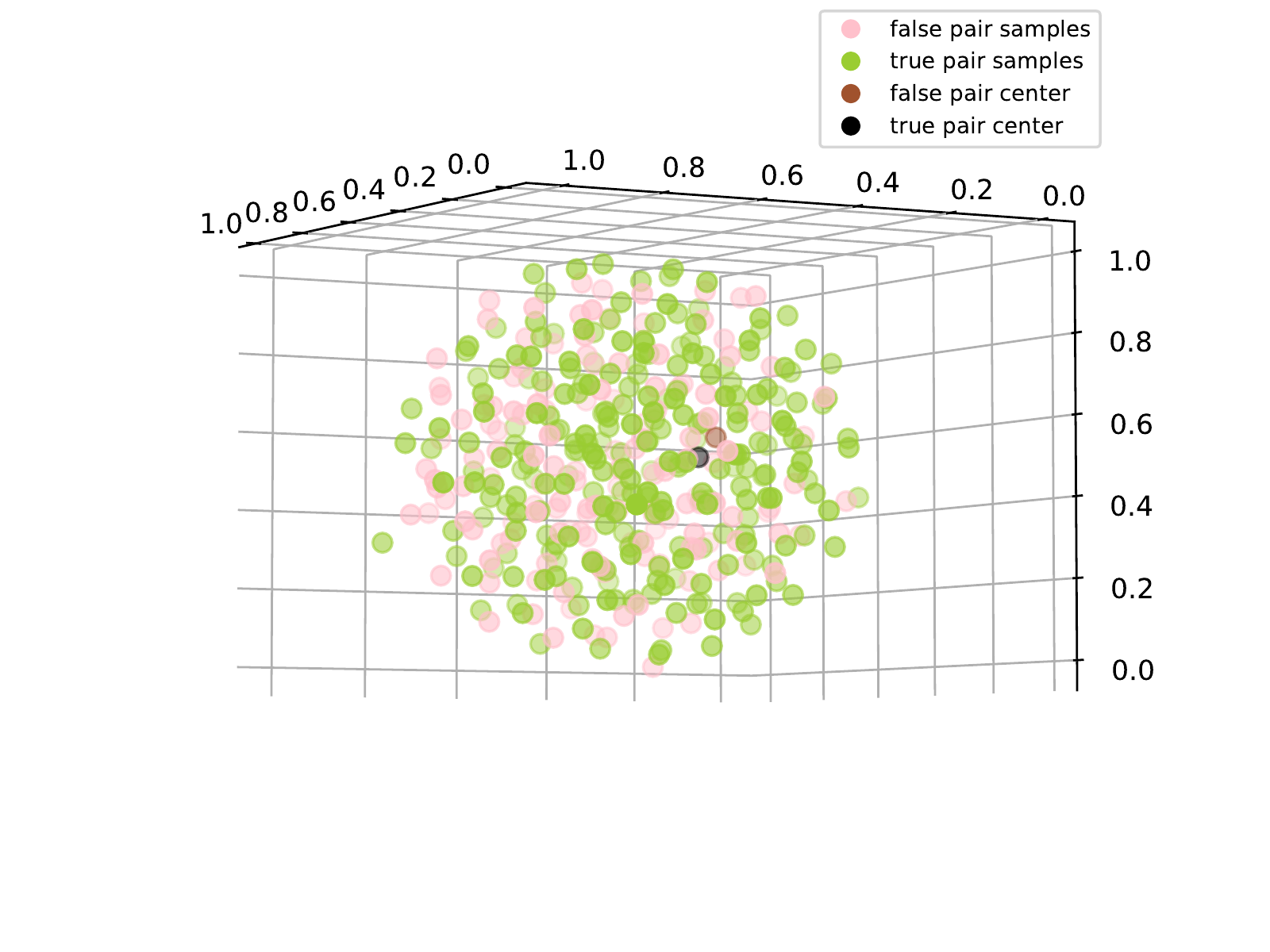}}
\subfigure[16res-before]{\includegraphics[width=0.244\textwidth]{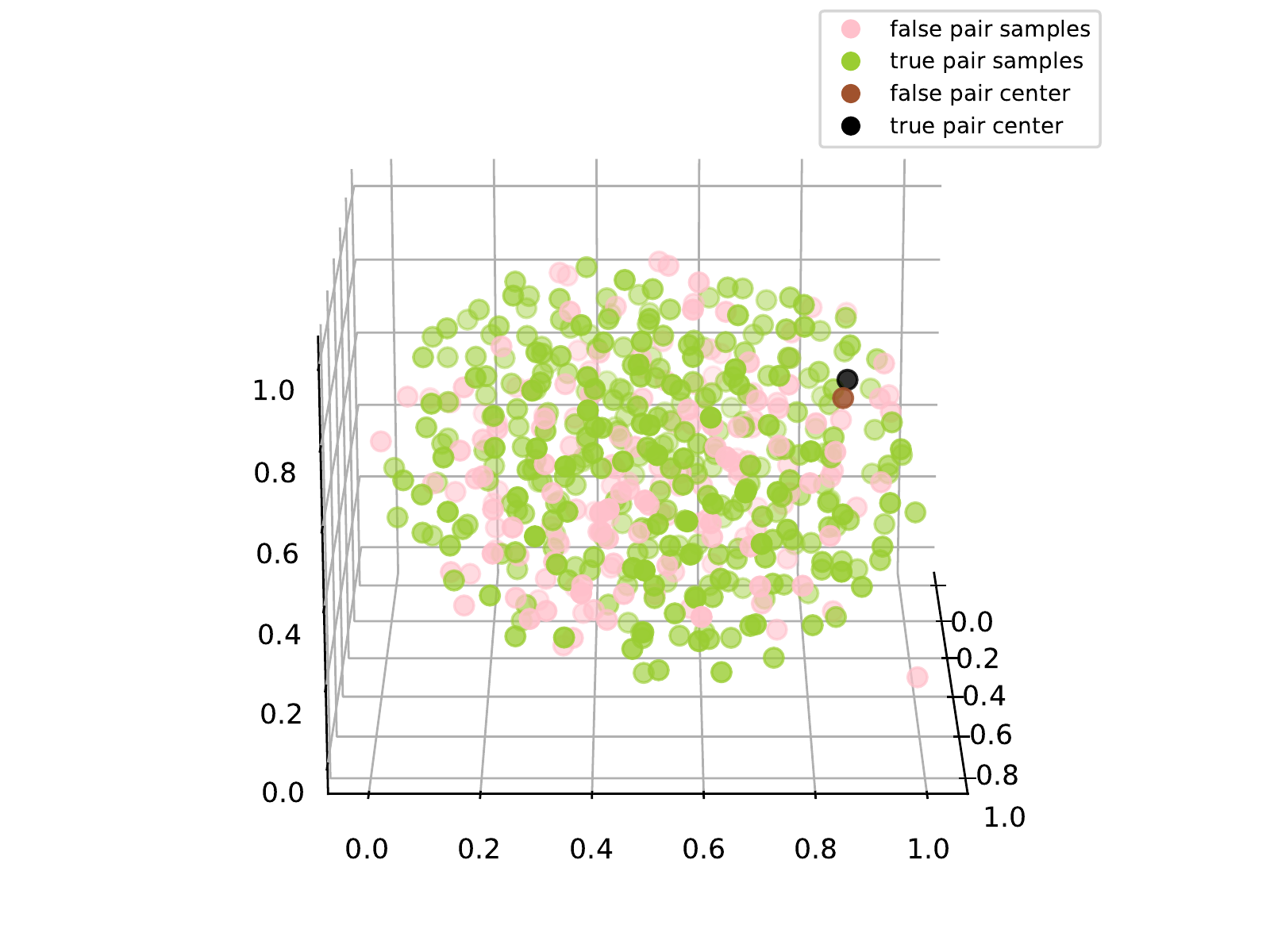}}
\\
\subfigure[14lap-after]{\includegraphics[width=0.244\textwidth]{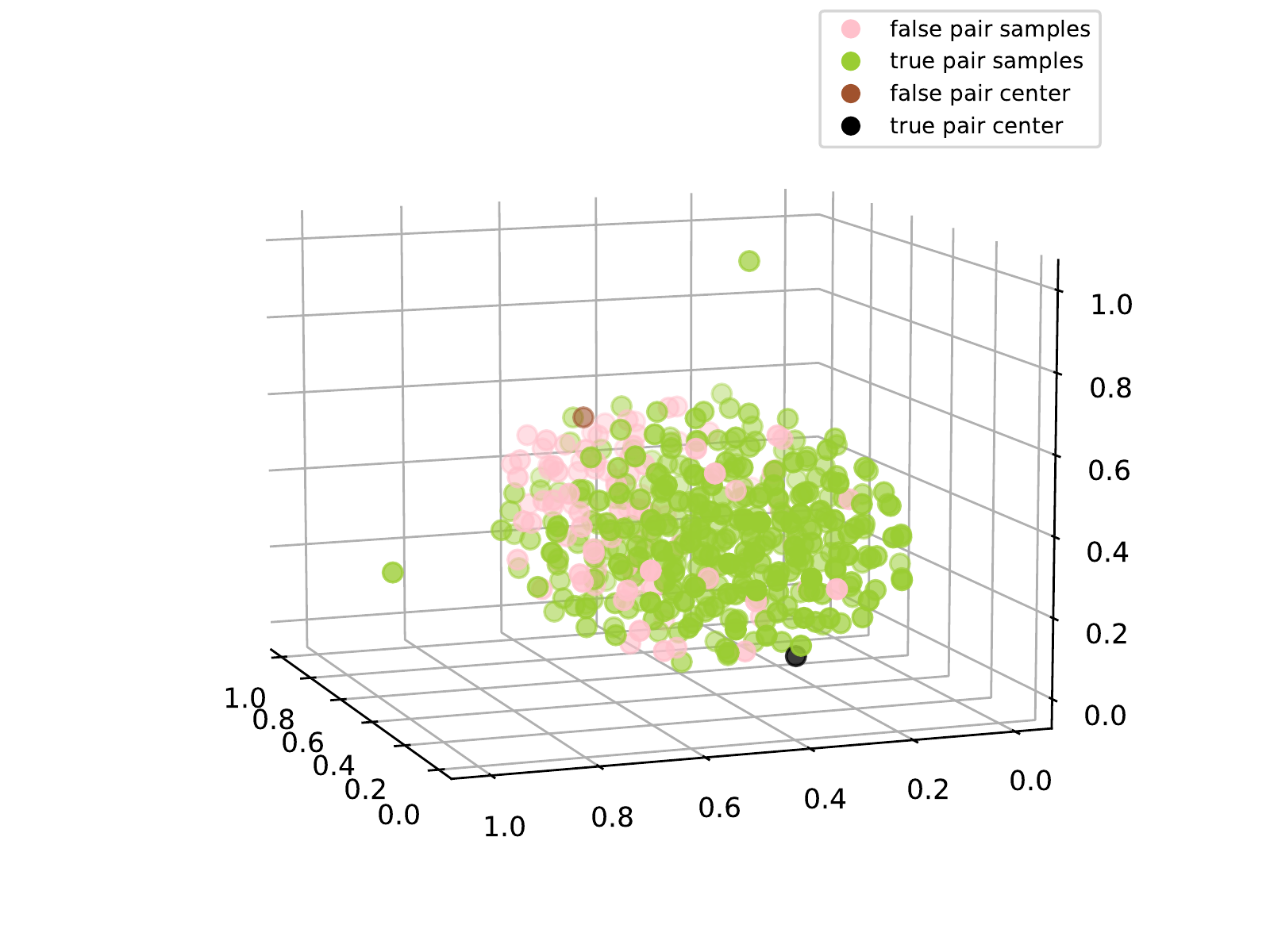}}
\subfigure[14res-after]{\includegraphics[width=0.244\textwidth]{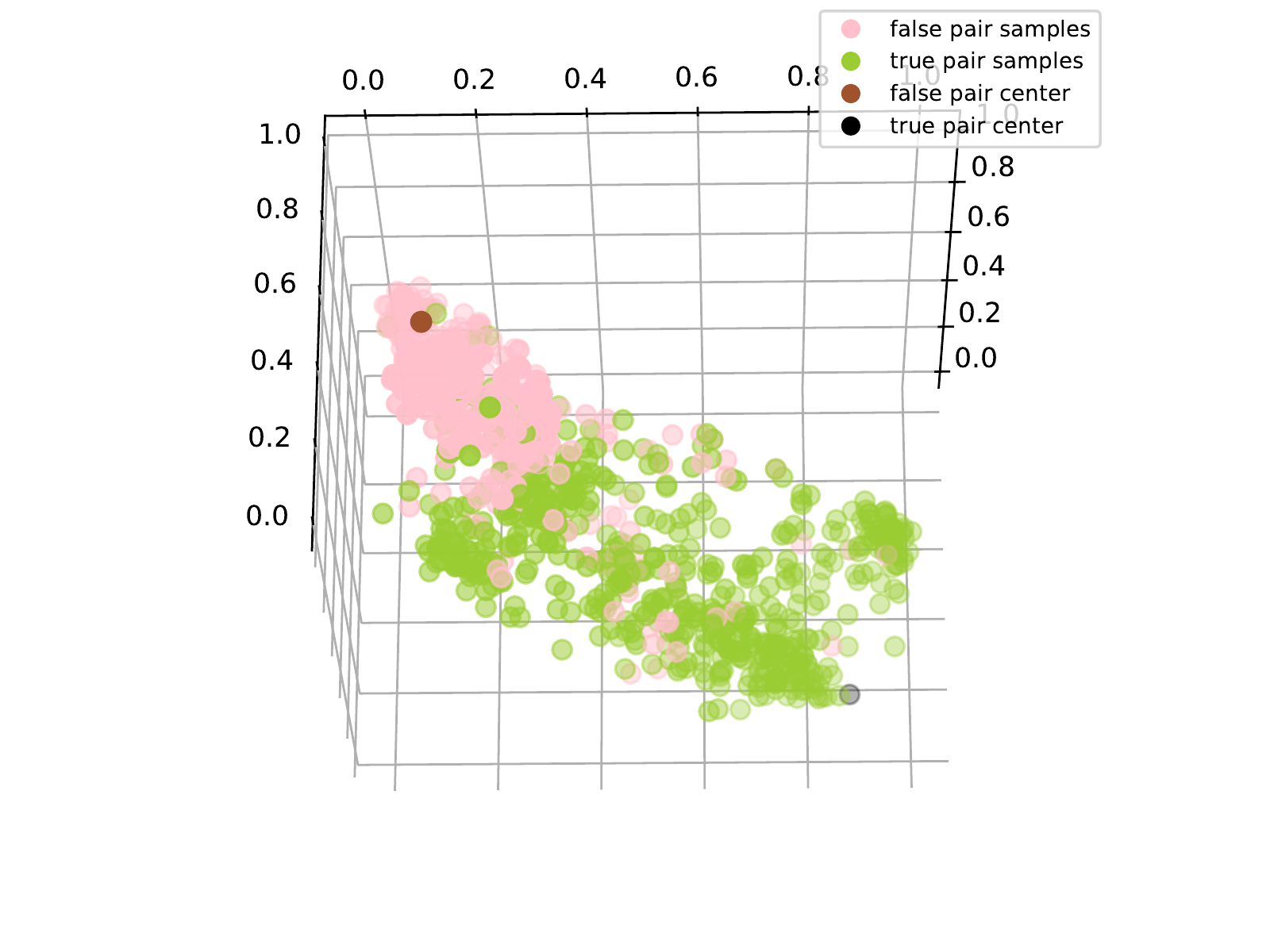}}
\subfigure[15res-after]{\includegraphics[width=0.244\textwidth]{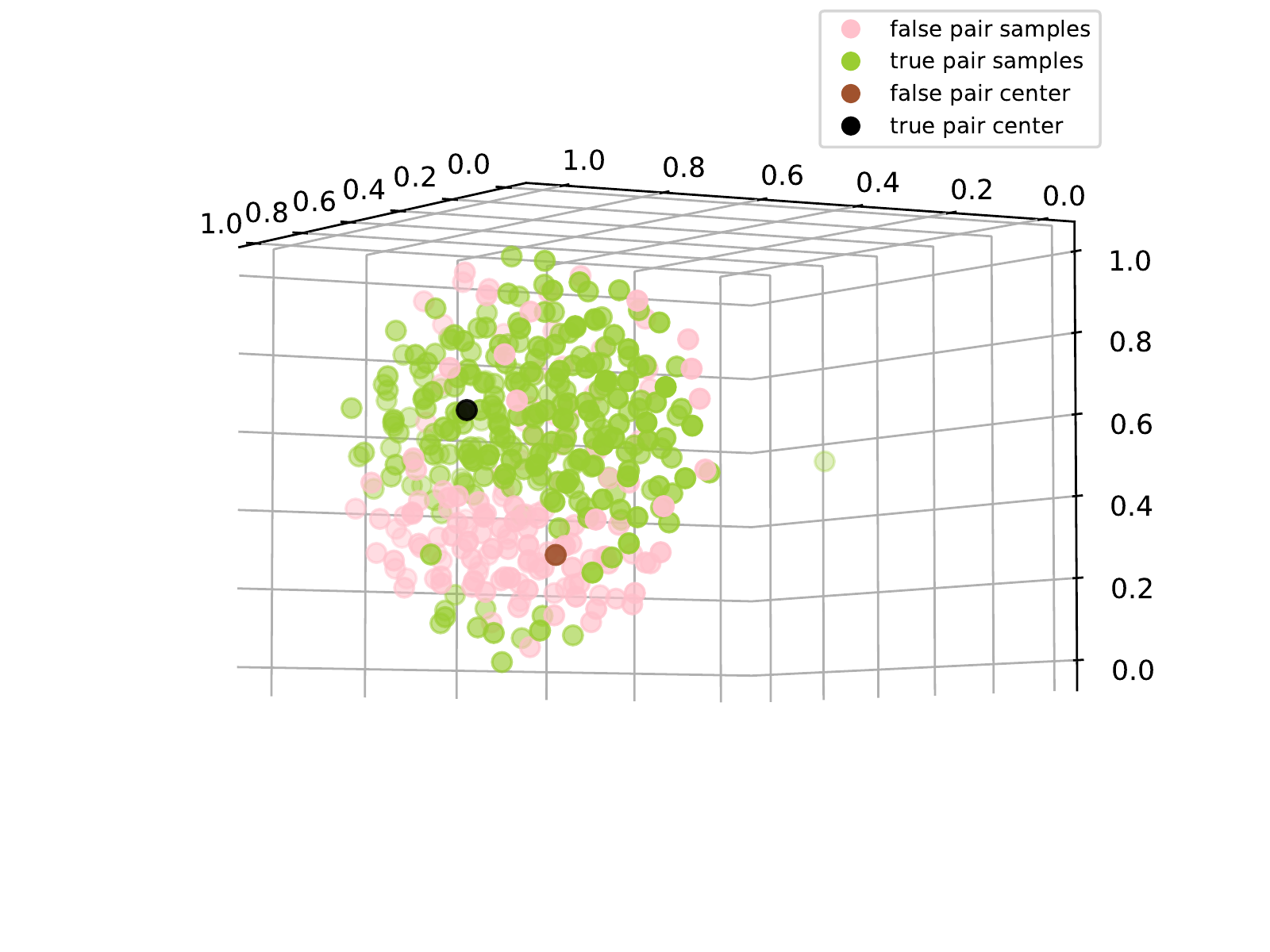}}
\subfigure[16res-after]{\includegraphics[width=0.244\textwidth]{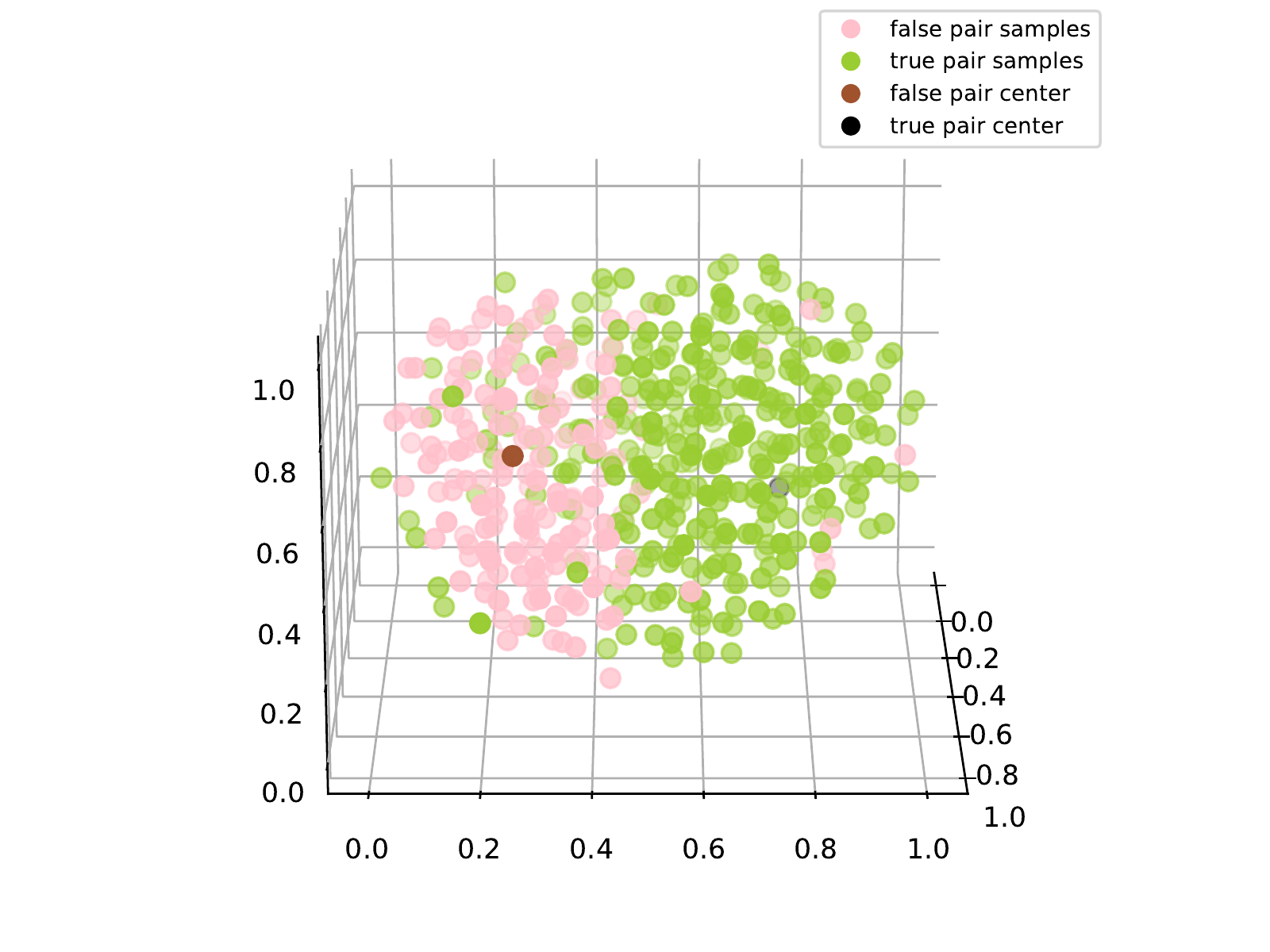}}
\caption{Visualization of pair representations of four datasets before and after training.}
\label{tsne}
\end{figure}

To offer a more comprehensible illustration of the impact of contrastive learning on the pair representations, we present Figure~\ref{tsne}, which shows the changes in pair representations before and after training on four datasets. 
In this figure, the black and red points represent the central points of true and false pairs, respectively. These points correspond to the vectors obtained from pair descriptions. The green and pink points represent true and false pair sample vectors from the input sequences.
Before training, the sample vectors are mixed and are difficult to distinguish with low discrimination. However, after training, they are separated into two distinct clusters and dispersed around their respective centers, forming a relatively clear boundary.
This boundary is determined by features learned from contrastive learning, which enables the vectors of centers to be similar with the corresponding positive term pairs, and dissimilar with the negative ones.
This result suggests that the introduction of contrastive learning allows different pair samples to learn the semantic features specific to their corresponding types.

\subsection{Case Study}
Table~\ref{tab5} presents the representative generated results by GAS and our approach. 
In the first example, the term ``crispiness" is used to describe ``batter", but GAS mistakenly matches this opinion term with ``prawns". 
In the second example, although GAS correctly extracts the aspect terms ``runs" and ``fans", it also extracts the non-opinion term ``huge bonus", resulting in incorrect matching relations. 
The first two examples indicate that our approach produces fewer pairing errors than the previous vanilla T5-based method.
In the third example, GAS generates redundant matches, while our approach do not suffer from this problem but has some errors in boundary detection. We believe that future work could attempt to correct these boundary errors.

\begin{table}[htbp]

  \centering

    \setlength\tabcolsep{10pt}
  \caption{Case study.}

  \resizebox{1.0\columnwidth}{!}{
    \begin{tabular}{>{\centering\arraybackslash}m{6cm}ccc}
    \hline
    \multicolumn{1}{c}{Review} & Ground-truth & GAS & Ours \\
    \hline
\multirow{4}{6cm}{\centering so about the prawns, they were fresh and had a slight crispiness about the batter … soooo good ... the walnuts were cut in smaller pieces and very crunchy and tasty.} & (prawns, fresh, positive) & (prawns, fresh, positive) & (prawns, fresh, positive) \\
& (batter, crispiness, positive) & \textcolor[rgb]{ 1,  0,  0}{(prawns, crispiness, positive)} & (batter, crispiness, positive) \\
& (walnuts, crunchy, positive) & (winters, crunchy, positive) & (walnuts, crunchy, positive) \\
& (walnuts, tasty, positive) & (winters, tasty, positive) & (walnuts, tasty, positive) \\
\hline
    
\multirow{4}{6cm}{\centering The Mac mini is about 8x smaller than my old computer which is a huge bonus and runs very quiet, actually the fans aren't audible unlike my old pc} & &  &  \\
& (runs, quiet, positive) & \textcolor[rgb]{ 1, 0, 0}{(runs, huge bonus, positive)} & (runs, quiet, positive) \\
& (fans, aren't audible, positive) & \textcolor[rgb]{ 1, 0, 0}{(fans, quiet, positive)} & (fans, aren't audible, positive) \\
& &  &  \\
\hline

\multirow{3}{6cm}{\centering BEST spicy tuna roll, great asian salad.} & (asian salad, great, positive) & (spicy tuna roll, BEST, positive) & \textcolor[rgb]{ 1, 0, 0}{(tuna roll, BEST, positive)} \\
& (spicy tuna roll, BEST, positive) & (spicy tuna roll, great, positive) & (asian salad, great, positive) \\
& / & (asian salad, great, positive) & / \\
     \hline
    \end{tabular}
    }
  \label{tab5}%
\end{table}%
\section{Conclusion}
\label{con}
In this paper, we propose a pairing information enhancement approach for Aspect Sentiment Triplet Extraction (ASTE), which incorporates contrastive learning during the training stage of the triplet extraction model.
We introduce contrastive learning objectives based on the pair description vectors and pair sample vectors to enhance the ability of model to pair aspect terms and opinion terms.
Experimental results demonstrate that our approach performs well on four ASTE datasets compared to several related classical and state-of-the-art triplet extraction methods.
Future directions include: incorporating additional pair knowledge; applications to other ABSA tasks.
\par
\par
\subsubsection{Acknowledgements.} The work is supported by National Nature Science Foundation of China (No.62176174).
\end{CJK*}
%
%
%
\end{sloppypar}

\bibliography{refs}
\bibliographystyle{splncs04}

\end{document}